\documentclass[11pt]{article}

\usepackage[margin=2.5cm]{geometry}
\usepackage{authblk}

\usepackage[T1]{fontenc}
\usepackage[utf8]{inputenc} % optional on modern LaTeX, kept for compatibility
\usepackage[bitstream-charter]{mathdesign} % Charter-like text + matching math

\usepackage{graphicx}
\usepackage{float}
\usepackage{multicol}

\usepackage{enumitem}
\usepackage{csquotes}

\usepackage{amsmath}
\usepackage{tikz}
\usetikzlibrary{calc,positioning,decorations.pathreplacing}
\usepackage{setspace}
\usepackage[dvipsnames]{xcolor}
\definecolor{custom-yellow}{HTML}{ffea94}
\definecolor{custom-yellow-darker}{HTML}{dbaf00}
\definecolor{custom-green}{HTML}{007849}
\definecolor{custom-blue}{HTML}{0375B4}
\definecolor{custom-blue-darker}{HTML}{035582}
\definecolor{custom-blue-lightest}{HTML}{cdedfe}

\usepackage{xurl} 
\usepackage{hyperref}
\hypersetup{
  colorlinks=true,
  allcolors=black,
  breaklinks=true
}

\usepackage{cleveref}

\usepackage{tocloft}
 
\usepackage{titlesec}
\renewcommand\thesection{\arabic{section}}
\titleformat{\section}
  {\huge}{\thesection}{2em}{}
\titleformat{\subsection}
  {\Large}{\thesubsection}{1em}{}

\usepackage[
  backend=biber,
  style=apa,
  natbib=true,
  autocite=inline, % footnote
]{biblatex}

\AtEveryBibitem{\clearfield{urldate}}

\newif\ifciteaudit
\citeauditfalse

\AtEveryCite{%
  \ifciteaudit
    \color{Magenta}\bfseries
  \fi
}

\newcommand{\CiteAuditOff}{\citeauditfalse}
\begin{document}
\CiteAuditOff

\title{Demonstrating Restraint}
\setcounter{footnote}{1}
\author[1]{L. C. R. Patell\thanks{\texttt{liam.patell@governance.ai}}}
\author[2]{O. E. Guest\thanks{\texttt{oliver@iaps.ai}}}
\affil[1]{GovAI}
\affil[2]{Institute for AI Policy and Strategy}
\maketitle
\thispagestyle{empty}

\begin{abstract}
Some have claimed that the future development of powerful AI systems would enable the United States to shift the international balance of power dramatically in its favor. Such a feat may not be technically possible; even so, if American AI development is \emph{perceived} as a sufficiently severe threat by its nation-state adversaries, then the risk that they take extreme preventive action against the United States may rise. To bolster its security against preventive action, the United States could aim to pursue a strategy of \emph{restraint} by demonstrating that it would not use powerful AI to threaten the survival of other nations. Drawing from the international relations literature that explores how states can make credible commitments, we sketch a set of options that the United States could employ to implement this strategy. In the most challenging setting, where it is certain that the US will unilaterally obtain powerful new capabilities, it is difficult to credibly commit to restraint, though an approach that layers significant policy effort with technical breakthroughs may make credibility achievable. If an adversary has realistic levels of uncertainty about the capabilities and intentions of the United States, a strategy of restraint becomes more feasible. Though restraint faces difficulties, it deserves to be weighed against alternative strategies that have been proposed for avoiding conflict during the transition to a world with advanced AI.
\end{abstract}

\newpage
{\begin{spacing}{0.9}
\pagenumbering{gobble}
\tableofcontents
\end{spacing}}
\newpage
\pagenumbering{arabic}

\section{ Introduction\label{introduction}}

Several AI developers in the United States are explicitly aiming to develop AI systems that outperform humans across most or all cognitive tasks \autocite{P-Amodei2024x,P-OpenAI2023o,P-Altman2025x,P-Zuckerberg2025o}. We will refer to such systems as \emph{powerful AI}.\footnote{Similar terms that are commonly used elsewhere include \emph{artificial general intelligence (AGI)} and \emph{superintelligence}, with the latter generally connotating a particularly high level of performance.} Some actors in other countries are also interested in developing powerful AI \autocite{P-Allen2025s,P-Schneider2025l}. Nevertheless, there is some chance that the U.S. would acquire powerful AI significantly before other countries do: the U.S. has several advantages in AI development, and there may be compounding effects, whereby a small initial lead in developing more capable AIs would grow into a much larger lead \autocite[][pp.~30--31]{P-Aarne2025u}.

Unilateral possession of powerful AI would have significant geopolitical implications. It might enable the U.S. to design offensive or defensive ``wonder weapons'', conduct devastating cyber-attacks, or engage in foreign interference through immaculately designed influence operations. Some have even claimed that unilateral possession of powerful AI might disrupt the nuclear deterrence equilibrium and yield a \emph{decisive strategic advantage (DSA)}: the ability to secure
its preferred outcomes across all levers of power and to sustain that position indefinitely \citep{P-Aschenbrenner2024v, P-Hendrycks2025p}.\footnote{We use the definition from \citep{CoatesForthcoming}. As we note below, Coates concludes that technical conditions render it unlikely that unilateral possession of powerful AI would yield a DSA.}

It is not at all clear that a DSA will be possible. The capabilities alluded to above are speculative. Additionally, achieving a DSA requires bringing the expected harm from a retaliatory attack below unacceptably disastrous levels; doing so would require a battery of demanding technical conditions to hold \citep{P-Mitre2025t,P-Winter-Levy2025m,CoatesForthcoming}.

Nevertheless, foreign governments might be extremely concerned by the potential strategic implications of powerful AI. The threat of a DSA may become increasingly salient to national security decision makers if AI progress marches on, even if the technical possibility of a DSA remains deeply uncertain. At the most extreme level, foreign governments might worry that the U.S. would use an AI-enabled DSA to undermine their country's very survival; we refer to this possibility as \emph{trampling}. If these threat perceptions are sufficiently strong, then U.S. adversaries may be tempted to engage in preventive conflict to prevent the U.S. from acquiring powerful AI and a potential DSA \citep{P-Powell2006u,P-Katzke2025n,P-Burdette2025p}.\footnote{Note that these authors vary dramatically in their assessment of the likelihood of preventive conflict, with Hendrycks et al. at the high probability extreme and Burdette et al. at the low probability extreme.} Adversaries could view the risk of escalation to nuclear war as a price worth paying to avoid an existential threat to their sovereignty.\footnote{Although not a focus of this paper, such a threat perception might also lead to other dangerous effects. In particular, foreign governments might engage in increasingly reckless AI development and deployment practices in an effort to keep pace with the US.}

\subsection{Introducing Restraint\label{sec:1.1-introducing-restraint}}

This paper explores one family of measures that U.S. policymakers could use to reduce the likelihood of preventive war: demonstrating \emph{restraint}. We use restraint to refer to a strategy where the U.S. would decide not to pose an \emph{existential} threat to its adversaries, \emph{even if the U.S. were to come to possess AI systems that by default would allow it to easily trample them.}\footnote{Throughout this paper we refer to existential threats \emph{to specific countries}. We note that such threats might differ from existential threats as they are sometimes discussed in AI discourse, which are to humanity as a whole. See for example \cite{P-CenterForAISafety2023h}.} This definition leaves out of scope various other measures to reduce the likelihood of preventive war caused by fear of an AI-enabled DSA. For example, one could also avoid developing a DSA in the first place, such as by deliberately preserving an international balance of AI capabilities \citep{P-Hendrycks2025p} or by refraining from developing superintelligence entirely \citep{P-Katzke2025n}. Restraint and related concepts have already been discussed in shorter outputs from researchers at CSIS and RAND \citep{P-Frelinger2026,P-Lonergan2026,P-Mazarr2026}.

We do not claim that restraint is necessarily the most promising option -- only that it is sufficiently plausible to be weighed against other options for reducing the risk of war from strategic competition.\footnote{In contrast, \citet{P-Hendrycks2025q}, for example, claim that ``states cannot trust that rivals won't use ASI against them'', without considering the possibility of strategically cultivating trust by demonstrating restraint.}

There are important limits to interventions based around restraint. If U.S. adversaries think that America is on track to acquire an AI-enabled DSA, and that the U.S. would want to use such a DSA to trample its adversaries, then the mechanisms described below would likely be insufficient to provide complete reassurance.

Those assumptions, however, might be \emph{much} more demanding than what will actually happen. When we relax those assumptions, we see that demonstrating restraint may be more tractable than it might first appear. Adversaries might not be certain that:

\begin{enumerate}
\item
  AI-enabled DSAs are technically possible;
\item
  The U.S. is on track to acquire one, even if they are possible;
\item
  The U.S. would want to use a DSA to trample them.
\end{enumerate}

Under such conditions, the amount the U.S. would need to do to credibly demonstrate restraint diminishes. Merely increasing adversaries' confidence that they would not be trampled might be sufficient to keep them below the threshold of escalating to preventive war.

Moreover, adversaries might believe that a U.S. DSA would only be temporary. For example, adversaries might believe that they would eventually catch up to the U.S. in AI capabilities or that they would acquire other strategic advantages outside of frontier AI development. This means that adversaries might not need to be confident that restraint measures would endure indefinitely -- only long enough for the strategic situation to change.

\subsection{Structure\label{sec:1.2-structure}}

The remaining sections of this paper explore restraint in more detail. \Cref{sec:2-theoretical-overview-of-restraint-measures} provides a theoretical overview of two ways in which the U.S. could credibly commit to restraint: costly signaling and a concept that we call \emph{foreclosure guarantees}. Costly signals convey information or shape U.S. incentives through the manipulation of different kinds of cost. \emph{Foreclosure guarantees} are actions that the U.S. could take that would render trampling adversaries extremely difficult, or even impossible. \Cref{sec:3-a-sketch-of-implementing-restraint-measures} presents a sketch of restraint measures that the U.S. could start to implement, focusing on measures that would not require major technical breakthroughs. \Cref{sec:4-speculative-mechanisms} discusses mechanisms that might be more effective for demonstrating restraint but that are particularly speculative, in part because they depend upon hypothetical future technologies. \Cref{sec:5-future-research} identifies various directions for further research. The sections are intended to be modular; we encourage readers to skip to the sections that are most relevant to them.

\subsection{Addressing Two Cross-Cutting Objections\label{sec:1.3-addressing-two-cross-cutting-objections}}

Before proceeding, we address two potential objections that apply to the paper as a whole.

\textbf{Which actors are relevant?} An initial objection is that this paper focuses too strongly on the U.S. \emph{government}. The trampling actions we described above would presumably be undertaken by parts of the government (such as the military), and the interventions we outline below mostly target the government. This focus may seem incongruent with the fact that, in the US, it is currently \emph{companies} that seem to have the strongest motivation and ability to develop powerful AI.

This discrepancy is less meaningful than it might seem. First, the U.S. government might procure highly strategic capabilities from U.S. companies, even though those companies remain independent from the government. The government has already demonstrated interest in procuring AI capabilities for military uses, as evidenced by the recent awards to Anthropic, Google, OpenAI, and xAI with a ceiling of US\$200 million \citep{P-ChiefDigitalAndArtificialIntelligenceOffice2025x}.\footnote{More generally, many military systems are made for the U.S. government primarily by defense contractors, not by the government itself.} Second, the U.S. government might become more closely intertwined with AI companies in the future \autocite{P-Aschenbrenner2024v,P-Harris2025k,P-Anderson-Samways2025e}. Third, even if there are, in fact, limited links between the U.S. government and U.S. companies that develop powerful AI, U.S. adversaries might incorrectly \emph{believe} that there are close links. Consider, for example, AlphaGo, a system developed by Google DeepMind\footnote{The organization was called DeepMind in 2017, and has since been renamed to Google DeepMind. Both then and now though, the organization was part of Alphabet and so clearly part of a company.} that achieved superhuman performance at the game of Go: ``Chinese military researchers consistently drew connections between Google DeepMind's capabilities and broader U.S. state power, illustrating how they viewed private firms' capabilities as an extension of state capabilities'' \citep{P-Xiao2025x}.

\textbf{Is restraint an undue concession?} A second -- more normative -- potential objection is that it is undesirable for the U.S. to make concessions to adversaries, which includes reassuring them about its intentions not to trample them.

This objection largely misses the mark: the motivation for restraint is not that it benefits adversaries but rather that it benefits the US. Preventive war with a peer or near-peer competitor would likely be extremely costly for the US, even if the U.S. would stand a good chance of winning. Restraint aims to benefit the U.S. by making such a war less likely. This strategy demands merely that the risk of preventive war is not a price worth paying to preserve the ability to trample these adversaries. There is precedent for this sort of decision-making; throughout the Cold War, there were points where the U.S. deliberately structured its nuclear forces to optimize for strategic stability rather than the capacity to win a nuclear war against the Soviet Union \autocite{P-Payne2022j}.

We turn now to a theoretical overview of restraint measures, drawn from the international relations (IR) literature, applied to the context of interstate competition and powerful AI.

\section{Theoretical Overview of Restraint Measures\label{sec:2-theoretical-overview-of-restraint-measures}}

Restraint depends upon being able to make credible commitments. A commitment is a pledge to carry out (or not carry out) an action in the future; a \emph{credible} commitment is one that adversaries believe will be upheld.\footnote{In game theory, adversaries specifically have this belief because \emph{it would be rational to uphold the commitment at the relevant point in time} \citep[][pp.~14--15]{P-Selten1975t}. In practice, adversaries may reflectively endorse these beliefs even if the game-theoretic criterion fails to hold.} Specifically, the strategy of restraint under investigation would require the U.S. to make credible commitments not to trample its adversaries, even if the U.S. acquired the capability to do so through an AI-enabled DSA.

Below, we discuss two \emph{approaches} to making credible commitments to restraint. First, the U.S. could aim to convey information to adversaries -- in a way that those adversaries would believe -- that the U.S. would not want to trample adversaries, even if easily able to do so. Second, the U.S. could \emph{shape the choice space} so that the U.S. would not be incentivized, or perhaps even able, to trample adversaries upon acquiring a DSA.\footnote{If the first approach suffices, then the problem of preventive conflict is a purely informational problem -- one that would be solved if states had complete information about one another's motives and capabilities. To the extent that conflict may stem from an incentive to renege on commitments to restraint, then reshaping the choice space is necessary. Note, however, that a signal that solves the information problem often also reshapes the choice space by creating the possibility of reputational harm.} We discuss two classes of \emph{mechanisms} through which the U.S. could take one or both of these approaches:

\begin{itemize}
\item
  \textbf{Costly signals} convey information or shape U.S. incentives through the manipulation of different kinds of cost.
\item
  \textbf{Foreclosure guarantees} are actions that the U.S. could potentially take after which it would be difficult or impossible for the U.S. to trample adversaries. Foreclose guarantees would shape the option space by taking the option of trampling off the table.
\end{itemize}

\subsection{Costly Signals\label{sec:2.1-costly-signals}}

Costly signals convey information or shape U.S. incentives through the manipulation of different kinds of cost. An example from international relations is mobilizing troops in a crisis \autocite[][p.~70]{P-Fearon1997g}. Because it is costly to mobilize troops, mobilization signals resolve to other countries, i.e. provides some information to them that the country is willing to pay nontrivial costs to resolve the crisis in its preferred direction. Mobilizing troops also shapes incentives; it puts the country in a better position to win a violent conflict that emerges from the crisis.

The concept of powerful AI, and, correspondingly, the idea that a country might acquire a decisive strategic advantage, exacerbates two challenges for costly signaling. We lay out these problems -- \emph{power shift} and \emph{type shift} -- before providing an overview of four kinds of costly signaling.\footnote{This decomposition stems from \citet{P-Quek2021u}, cited by \citet{P-Imbrie2023x}, and constitutes a logically valid decomposition of the space of possible costly signals.} These different kinds of signaling are likely to vary in their effectiveness in the context of a DSA. Costs that have purely informational effects will struggle to separate restrained actors from aggressive ones. This is because the gains to an aggressive actor from exploiting a DSA might be so large that an aggressive actor would be willing to bide their time until in a position to trample. Costs that alter the incentive structure are more structurally promising, but they are technically difficult to implement and the size of the costs required risk straining political feasibility.

\subsubsection{Challenges to Costly Signaling\label{sec:2.1.1-challenges-to-costly-signaling}}

The concept of powerful AI, and corresponding DSA, exacerbates at least two challenges to costly signaling. These are the \emph{power shift} and \emph{type shift} problems.

\textbf{Power shift problem:} It would not be sufficient for signaling mechanisms to convey information about the U.S. in the status quo or shape the incentives that the U.S. currently faces.\footnote{\citet{P-Powell2006u} provides a canonical example of how power shift can lead bargaining to collapse into conflict even when they possess complete information.} Successful mechanisms would also need to do so for a potential future where the U.S. has acquired powerful AI systems and a corresponding DSA. This feat might be significantly more challenging to perform successfully. For example, information about how the U.S. currently acts might not be relevant to a scenario where the U.S. is dramatically less constrained by the international system because AI has given the U.S. so much more power, relative to other countries. Likewise, incentives that might seem substantial in the present day might be less substantial in the future. For example, if powerful AI can bring about massive economic growth, then absolute amounts of money at stake that are very large by today's standards might seem trivial by tomorrow's standards.

\textbf{Type shift problem:} In canonical explorations of costly signals in the IR literature, states have a ``type'', or set of unobserved characteristics.\footnote{\citet{P-Fearon1994b, P-Fearon1995x, P-Fearon1997g}; \citet[p.~153]{P-Fey2011y}.} Here, the relevant characteristic is the degree to which the U.S. would want to exploit a DSA to trample rivals, should it achieve a DSA. A core assumption of the formal solutions to these games is that type does not change over time. Yet, in reality, type can oscillate wildly -- both because the personnel that make decisions about countries' actions can change and because the aims of individual personnel can evolve over time. As a result, a signal that would be credible if type were fixed might \emph{fail} to be credible given that type can change.\footnote{Appendix F demonstrates this problem formally.} This problem is related to the adage that absolute power corrupts absolutely: an adversary may believe that U.S. preferences will veer toward aggression as dominance becomes an increasingly achievable possibility. Because the degree of power conferred by a DSA would lie far outside the distribution of circumstances that government personnel have been exposed to, adversaries may be more likely to assume power-seeking intentions by default.

These two problems are complementary. The power shift problem primarily concerns the magnitude of costs: even substantial commitments may be overwhelmed by the prospective gains from exploiting a DSA. The type shift problem concerns the stability of intentions: even if adversaries believe current U.S. preferences are benign, they may doubt that those preferences will persist once overwhelming power becomes attainable. A credible commitment to restraint must address both challenges.

\subsubsection{Tying Hands\label{sec:2.1.2-tying-hands}}

Tying hands increases the cost of reneging on a commitment in the future.\footnote{In other words, they are \emph{ex post} -- they are incurred in the future -- and \emph{contingent} -- they depend on the signaler's actions.} If the commitment is upheld, then the cost does not occur \citep{P-Fearon1997g}.\footnote{In the original presentation, Fearon defines tying hands signaling as making \emph{backing down} from conflict costly. Because we investigate credible assurances, rather than credible threats -- restraint, rather than resolve -- we define tying hands as making trampling your adversaries more costly.} In other words, these signals involve ``leveraging future pay-offs as collateral to incentivize a commitment in the present'' \citep{P-Dafoe2014z}.

A prototypical example of tying hands is making public commitments so that domestic and international audiences would hold the U.S. government accountable if it reneges. The government could, for example, publicly proclaim that it has no intention of impugning the sovereignty of its adversaries and seeks no quarrel with foes across an ocean. In accordance with that proclamation, the government could vow to use powerful AI -- should it be obtained -- only in ways that do not undermine others' sovereignty.

The credibility of a tying hands signal can depend on the reputation of its originator: a reputation for restraint increases the credibility of reassurance \citep{P-Cebul2021f}. This operates through at least two channels. First, an adversary may simply be predisposed to believe the commitments of states (or individual government personnel) with a track record of restraint.\footnote{Though note that this mechanism does not constitute tying hands \emph{per se}.} Second, the prospect of audience costs -- reputational damage from breaking one's word -- may change the calculus of the choice situation.\footnote{\citet{P-Takei2025o} concludes from a survey experiment that audience costs can serve to ensure the credibility of assurances.} If America values its reputation highly enough, an adversary might believe that America would not want to incur the reputational hit from reneging on a commitment; as a result, that commitment could be seen as credible.

The power shift problem is particularly acute here. For a tying hands signal to credibly reassure, it must impose a cost larger than the gains even a highly ``aggressive'' America would perceive from exploiting a DSA against its adversaries.\footnote{See Appendix C for a worked example of this solution to preventive conflict.} Otherwise, an aggressive actor could feign restraint, wait to obtain a DSA, and then renege. Because the potential gains from exploitation could be perceived as enormous, credible reassurance may require costs that are much larger than typical audience costs.

Audience costs face additional challenges in the DSA context. Domestically, it is unclear that American voters would want to punish their leaders for breaking promises to foreign adversaries -- particularly if doing so appeared to benefit the United States. Internationally, states typically hold one another accountable through repeat interactions, which presupposes that other states will have future opportunities to impose consequences. But if a state truly possessed AI that could enable a DSA, then it might outgrow the international system entirely, achieve dominance, or simply care less about how other states interact with it. In such cases, international audience costs could become negligible.

That said, audience costs are unlikely to be zero. The opportunity cost of destroying one's reputation -- and having to rule as a tyrannical global hegemon rather than as a beacon of prosperity -- may be significant compared to trajectories where the United States simply uses powerful AI to outgrow and out-innovate its adversaries. But this residual cost is probably insufficient to make tying hands signals credible on their own.

\subsubsection{Sunk Costs\label{sec:2.1.3-sunk-costs}}

Sunk costs are costs that are incurred immediately, no matter what later actions occur.\footnote{In other words, they are \emph{ex ante} and \emph{noncontingent}.} As a result, they do not affect the \emph{relative} value of different actions \citep{P-Fearon1997g}. Instead, the fact that the signaler is willing to pay the sunk cost can reveal information about their preferences and intentions.\footnote{See Appendix A for an adaptation of \citet{P-Fearon1997g}'s model to the DSA context.} A canonical example from \citet{P-Fearon1997g} is mobilizing troops, which incurs an immediate cost while also changing the expected value of conflict.\footnote{The expected value of conflict plausibly changes because mobilizing troops early might make a country more likely to win any conflict that does occur.} The credibility effect of this signal can be decomposed into two components: the sunk cost of expending resources in the present, as well as the tying hands cost that shapes the choice space in the future.

Sunk costs are poorly suited to demonstrating restraint in contexts where states might acquire a DSA.\footnote{The efficacy of sunk costs for signaling restraint draws out differences between restraint that are credible despite a DSA and \citet{P-Fearon1997g}'s canonical model of credible threats.} States on a trajectory to obtain a DSA would benefit from signaling restraint, regardless of whether the state is in fact restrained or aggressive; in either case, the state would likely be keen to avoid its adversaries launching a preventive war. For sunk costs to succeed at signaling restraint, restrained states must be more motivated to signal restraint than aggressive states would be, and so more willing to pay sunk costs. But aggressive states \emph{have more to gain} from avoiding preventive conflict than do restrained states: they gain an opportunity to exert unparalleled dominance and trample their adversaries. The power shift problem bites here: \emph{both types} incur no significant cost if they choose to exploit a DSA. Hence, because sunk costs accrue \emph{no matter what later actions occur,} they do not change what a state prefers to do once it gets a DSA. For that reason, an aggressive actor is \emph{always} incentivized to mimic the sunk costs, feign restraint, and bide its time before trampling its adversaries.\footnote{In more technical terms, sunk costs signals can create a credible commitment only if there can be a separating equilibrium. The key difference between modeling a DSA and canonical signaling models is that one can nigh-costlessly make credible threats once one has obtained a DSA: by definition, such an advantage gives an adversary a vanishingly small chance of emerging victorious from a conflict. This practically risk-free conflict creates a structural difference that erases the possibility of a separating equilibrium.} Despite these difficulties, however, tying-hands-based reputational costs might be \emph{bolstered} by signaling in this way: domestic audiences may dislike burning resources more than reneging alone.

\subsubsection{Installment Costs\label{sec:2.1.4-installment-costs}}

Installment costs are costs that will be incurred in the future, no matter what actions the signaler takes.\footnote{They are \emph{ex post} and \emph{noncontingent}.} They can be incurred in a single installment at some point in the future, or they can be paid as a stream of continuous installments \citep{P-Quek2021u}. Quek provides the example of building a nuclear base in a foreign country to signal resolve to defend that country: ``the ex-ante costs of building the base are sunk costs; the ex-post costs of maintaining the nuclear base are installment costs'' (ibid.).

Installment costs face a similar problem to sunk costs as a mechanism for signaling restraint.\footnote{See Appendix D for installment costs in the context of a DSA.} Because installment costs are incurred no matter what future actions the signaler takes, an aggressive actor will be willing to pay to feign restraint in order to exploit later on.

\subsubsection{Reducible Costs\label{sec:2.1.5-reducible-costs}}

Reducible costs are paid up front but can be offset in the future depending on the actions of the signaler \citep{P-Quek2021u}.\footnote{See Appendix E for a model of reducible costs.} Quek provides the example of building bomb shelters for civilians: doing so incurs an immediate cost but decreases the cost of war in the future. If the state follows through on its threat and goes to war, it recoups some of the initial expenditure through reduced wartime losses.

Quek's example involves signaling resolve -- the state recovers costs by \emph{fighting}. This paper adapts the concept for signaling restraint: the state would recover costs by \emph{not} exploiting a DSA. For instance, \cref{sec:4.1-escrow-mechanisms} discusses placing resources in escrow, where the U.S. would forfeit assets if it violated a commitment to restraint but recover them if it honored the commitment.

Unlike tying hands signals, which rely on reputational costs, reducible costs create concrete material incentives for restraint. This sidesteps several problems identified above: reducible costs do not depend on domestic audiences wanting to punish leaders for breaking promises to adversaries, nor do they assume that international audiences will retain leverage over a state that has obtained a DSA.

Whether reducible costs can credibly signal restraint, however, depends on whether the costs can be scaled to outweigh the gains from exploitation. The power shift problem looms here as elsewhere: because the value of acting on a DSA could be enormous, even substantial material costs may be insufficient to change the calculus for an aggressive actor.

A further limitation is novelty. Implementations of reducible costs for restraint -- as opposed to resolve -- are unfamiliar and relatively unprecedented. Escrow mechanisms, for example, lie well outside the standard policy repertoire. Nevertheless, reducible costs remain among the more promising costly signaling mechanisms. \Cref{sec:4.1-escrow-mechanisms,sec:4.3-automatic-degradation} explore their implementation in greater detail.

\subsection{Foreclosure Guarantees\label{sec:2.2-foreclosure-guarantees}}

More speculatively, the U.S. could potentially take actions in the present that would strategically foreclose the possibility of taking certain actions in the future.\footnote{In this respect, foreclosure guarantees can be seen as particularly strong versions of tying hands signals that embody Schelling's ``paradox'' that ``the power to constrain an adversary may depend on the power to bind oneself'' \citep{P-Schelling1981a}.} Specifically, the U.S. would take actions that would make it impossible for the U.S. to trample adversaries in the future, even if the U.S. were in a position that would otherwise grant it an AI-enabled DSA. If adversaries were confident that the U.S. had implemented foreclosure guarantees, then they would be completely reassured that the U.S. would not trample them in future.

One potentially promising type of foreclosure guarantee concerns making guarantees about either the capabilities or propensities of (future) AI systems.\footnote{This taxonomy is taken from \citet{P-Shevlane2023a}.}

\begin{itemize}
\item
  \textbf{Capabilities:} Guarantee that AI systems would not have capabilities that would be likely to lead to a DSA. Returning to the examples earlier in the paper, this might mean refraining from capabilities to design wonder weapons, to conduct devastating cyber-attacks, or to carry out immaculately designed foreign influence operations. Implementing such a guarantee would likely require deliberate effort: as AI systems become more generally capable, DSA-relevant capabilities may emerge as a natural byproduct of that general capability growth.\footnote{See, for example, point two in \citet{P-Bowman2023k}.} Guaranteeing their absence would therefore mean actively removing or suppressing capabilities that would otherwise develop, rather than simply refraining from adding them.
\item
  \textbf{Propensities:} Guarantee that AI systems would not use the capabilities described in the previous bullet, even if they had them \emph{and if the operator requested them.} For example, AI systems with such a guarantee would reliably refuse to assist with requests to design wonder weapons.
\end{itemize}

There are several significant difficulties associated with foreclosure guarantees.

\textbf{Steering AI systems' behavior:} Making AI systems reliably refuse certain kinds of requests is an unsolved technical problem. There are many existing examples of users being able to elicit behaviors from AI systems that those systems' developers did not want users to be able to elicit \autocite{P-Wei2023y}. Consequently, even if making a good-faith effort, it would likely be difficult to build an AI system that has DSA-relevant capabilities but does not have the propensity to use them.

\textbf{Verification:} Even if the U.S. were to only build AI systems that lacked the capabilities and/or propensity to assist with trampling U.S. adversaries, it would be extremely difficult for adversaries to be confident of this fact. Adversaries would need to believe both of the following claims, both of which are difficult to verify.\footnote{These subgoals for AI verification are taken from \citet{P-Baker2025k}, as are the claims in the bullets about the difficulty of AI verification. See also \citet[][p.~3]{P-Harack2025v} on the difficulty of verifying negative claims about AI. An example of a negative claim would be, ``there are no AI systems with a propensity to trample''.}

\begin{itemize}
\item
  \emph{For highly capable AI systems that are known to the adversaries, these AI systems do not have capabilities and/or propensity to trample}. The verification technologies that would be needed to verify such a claim are relatively nascent. Verification would be particularly difficult if actors in the U.S. want to keep secret some information about their AI development, such as to avoid diffusing information about how to build more capable AI systems to U.S. adversaries.\footnote{A further difficulty with verification is that the techniques that are often used to ensure that AI systems refuse certain requests, such as RLHF and RLAIF, can be cheaply undone if one has access to the model weights \citep{P-Lermen2023i,P-Arditi2024r}. As a result, the U.S. might need to demonstrate not just that the original model has a certain propensity, but that this model has not since been changed to alter this propensity.}
\item
  \emph{There are no highly capable AI systems that are not known to the adversaries}. Verifying this claim might require relatively invasive mechanisms, such as tracking large amounts of compute within the U.S. to rule out secret data centers capable of training undisclosed, highly capable AI models.
\end{itemize}

\textbf{Dual-use AI inventions:} Powerful AI would presumably be useful for contributing to scientific discovery and R\&D. The resulting technologies could potentially be exploited by the U.S. to trample adversaries, without depending on the powerful AI itself. This dynamic implies that foreclosure guarantees might need to control not just whether AI systems themselves can be used to trample rivals but also whether they contribute to any other technologies that could be used for trampling. This scope significantly broadens the constraints that would need to be imposed on AI systems to provide foreclosure guarantees. Such broad constraints might be both technically challenging to implement, as well as prohibitively costly, since they might prevent AIs from assisting with beneficial but dual-use science and R\&D efforts.

\textbf{Succession:} AI systems are already helpful for developing future generations of AI systems, and some researchers have argued that this phenomenon will become more pronounced over time.\footnote{For example, \citet{P-Sett2024c}.} This dynamic poses a challenge for foreclosure guarantees: even if the U.S. can guarantee that some iteration of an AI system is unable or unwilling to assist with trampling, can the U.S. guarantee that the system will not assist with developing a successor AI system that \emph{does} assist? A possible solution is to guarantee that, if a system is assisting with AI development, it will only do so in a way that preserves certain guarantees. If this mechanism successfully iterates, then system $n$ would also refuse to help create system $n+1$ unless $n+1$ will include a guarantee.

This mechanism would likely be extremely technically challenging to implement. As noted above, it is difficult to reliably steer AI behavior. Steering behavior in a way that reliably persists across many generations of AI systems would presumably be even harder. And even if it were technically possible, developing this sort of mechanism would also carry major risks. The ability to modify the goals of AI systems is often thought of as a prerequisite for reducing accident risks from powerful AI;\footnote{See, for example, discussion of corrigibility and the safe shutdown problem in \citet{P-Soares2015b} and \citet{P-Thornley2024k}.} the mechanism we outline would likely involve deliberately suppressing that property. Additionally, the ability to ``lock in'' how AI systems will behave indefinitely is ripe for misuse; the same technology that could be used to lock in principles like state sovereignty could potentially be used to lock in poorly chosen or malicious principles.\footnote{For a speculative discussion of issues around AI and lock-in, see \citet{P-Finnveden2022f}.}

Having discussed what a commitment to restraint might look like at a theoretical level, we now turn to a sketch for how it could be implemented by the US.

\section{A Sketch of Implementing Restraint Measures\label{sec:3-a-sketch-of-implementing-restraint-measures}}

In this section, we articulate a package of measures that the U.S. could implement, if the U.S. wanted to credibly commit to restraint in the near future.\footnote{The focus on the near future means that we avoid discussing particularly speculative technologies. See \cref{sec:4-speculative-mechanisms} for discussion of such technologies.} These measures implement the approaches to restraint, i.e. costly signaling and foreclosure guarantees, that were discussed theoretically in the previous section. Although any individual measure would likely be inadequate, layering several of them may be sufficient to provide reassurance to adversaries that a commitment to restraint is credible. We focus on measures relating to diplomacy (\cref{sec:3.1-diplomacy}), interventions on AI systems (\cref{sec:3.2-interventions-on-ai-systems}), and institutional design (\cref{sec:3.3-institutional-design}).

\subsection{Diplomacy\label{sec:3.1-diplomacy}}

Diplomacy can contribute to a strategy of restraint in several ways that include both official government-to-government communications and Track II initiatives. In this section, we discuss how diplomacy might help to reassure adversaries; we then discuss its limitations.

First, diplomacy can provide adversaries with information that updates their beliefs about U.S. intentions. Foreign governments face genuine uncertainty about how the U.S. would use powerful AI. Diplomatic engagement, particularly sustained interaction over time, provides some evidence about U.S. priorities and motivations. Adversaries will be aware that diplomatic statements can be strategic, so this evidence is noisy, but it is not valueless. For example, if adversaries observe U.S. officials consistently emphasizing particular goals, these observations might shift probability mass toward more benign interpretations of U.S. intentions.

Second, diplomacy can highlight areas where U.S. and adversary interests overlap. Shared interests affect the overall expected value of U.S. AI development from an adversary's perspective. If the U.S. uses powerful AI to accelerate progress on disease eradication or other global challenges, adversaries also benefit from these advances. Diplomatic emphasis on such applications can make salient the dimensions along which U.S. AI progress is positive-sum, potentially offsetting concerns about coercion on other dimensions.

Third, diplomacy can serve as infrastructure for negotiating and implementing more substantive commitments. The diplomatic mechanisms discussed above are unlikely to be sufficient on their own to credibly demonstrate restraint. However, diplomacy provides the fora, relationships, and communication channels through which more costly signals can be negotiated and monitored.

Fourth, regular diplomatic interactions can help the U.S. gauge whether its commitments to restraint are being perceived as credible and whether preventive measures are under consideration. Diplomatic assurances may imperfectly predict the behavior of a state that has acquired overwhelming power, but \emph{ongoing} diplomacy can help with the type shift problem. Continuously engaging in diplomacy refreshes reassurances, thereby providing evidence against shifting aims.

Despite these potential contributions, diplomacy faces important limitations in the context of AI-enabled DSAs. Private diplomatic assurances are a canonical example of ``cheap talk'': statements that are costless to make and therefore provide limited evidence of genuine commitment. The limitations of cheap talk are particularly acute when the stakes are high, as they would be in scenarios involving potential DSAs. 

Additionally, even if diplomacy provides useful information about \emph{current} U.S. intentions, adversaries face a distinct uncertainty about how the U.S. would behave \emph{after} acquiring a DSA. Part of the difficulty of committing to restraint is that it might not be obvious, until it is too late, whether the commitments have been seen as sufficiently credible to convince an adversary to refrain from preventive war. If the atmosphere of diplomatic discourse sours, then stronger mechanisms may be required to alleviate tensions. For these reasons, diplomacy is best understood as a complement to the costlier mechanisms discussed in the remainder of this section, rather than as a standalone signal of restraint.

\subsection{Interventions on AI Systems\label{sec:3.2-interventions-on-ai-systems}}

As a second set of measures, the U.S. could intervene technically on AI systems, or the hardware underlying them, to reduce the propensity or capability of these systems to assist with trampling other countries. We discuss two mechanisms for achieving this goal: writing sovereignty into model specs and using hardware-enabled governance mechanisms to avoid AI systems with capabilities specifically relevant to trampling.

\subsubsection{Writing Sovereignty into Model Specs\label{sec:3.2.1-writing-sovereignty-into-model-specs}}

The U.S. has made public pledges to respect the sovereignty of other countries, such as in the UN Charter and the Shanghai Communiqué.\footnote{\cite[p.~229]{P-Thomson1995j}; \cite{P-AmericanExperience2019o}.} The U.S. could potentially intervene on U.S. AI systems to make them more likely to act in ways consistent with valuing sovereignty. Doing so could strengthen the U.S. commitment to restraint, since trampling would be a major violation of sovereignty.

One way to make AI systems more likely to act in accordance with sovereignty principles would be to write these principles into \textbf{model specifications} (hereafter, ``model specs''). Model specs are documents that define the behavior that a developer intends for an AI system to exhibit.\footnote{The term ``constitution'' is sometimes used for a similar concept. \citet{P-Anthropic2023e,ChanForthcoming}.} Various techniques, such as reinforcement learning from human feedback (RLHF), can be used to make AIs more likely to comply with the model spec, though these techniques sometimes fail to ensure that AIs behave as the developer intended.\footnote{\cites{P-OKeefe2025f}[point 4]{P-Bowman2023k}.} This proposal is similar to the idea of treaty-following AI.\footnote{\cite{P-Maas2025c}, see also \cite{P-OKeefe2025f}.} An AI that follows treaties that guarantee sovereignty, such as the UN Charter, is analogous to an AI that has a model spec that includes sovereignty principles.

Writing sovereignty into model specs would provide a signal that the U.S. would continue to value sovereignty, even if it were relatively easy for actors in the U.S. to circumvent these specifications. This signal could be strengthened by leveraging institutional processes to make it harder to remove sovereignty from these specs; we discuss institutional mechanisms in more detail below.

Much more ambitiously, one could try to use model specs to create guarantees that all U.S. AI systems would always respect sovereignty. Doing so would create a foreclosure guarantee; it would become impossible for the U.S. to use its AI systems to trample, since they would refuse to assist with that violation of sovereignty. However, as discussed in \cref{sec:2.2-foreclosure-guarantees}, such efforts face a litany of technical difficulties, and might not be desirable even if they were possible.

\subsubsection{Hardware Mechanisms to Avoid AIs with Certain Capabilities\label{sec:3.2.2-hardware-mechanisms-to-avoid-ais-with-certain-capabilities}}

A second way to intervene on AI systems would be to ensure that AI systems do not possess capabilities that are particularly relevant to trampling others, even if those AIs systems are otherwise highly capable. Flexible hardware-enabled guarantees (FlexHEGs), a class of mechanisms that operate on the hardware used to run AI systems, have been proposed as one way to enforce this sort of limitation \autocite[][p.~34]{P-Aarne2025u}. However, the technology required for FlexHEGs might still require ``several years'' to develop \autocite[][p.~16]{P-Petrie2025t}. Additionally, this mechanism faces the problems discussed in \cref{sec:2-theoretical-overview-of-restraint-measures} with refraining from developing certain particularly threatening capabilities. For example, even if powerful AI systems do not have clearly threatening capabilities, they might still be able to contribute to R\&D in general, and the results of this R\&D might themselves be useful for trampling adversaries.\footnote{Petrie and Aarne, two of the main FlexHEG authors, provide an additional detailed discussion of how the mechanisms could be used to preserve an international balance of power, even if one country would by default achieve AI capabilities progress much above its rivals \citep[][pp.~30-35]{P-Aarne2025u}. Similar to this paper, the motivation is on reducing risks of conflict associated with the specter of one country becoming much more powerful than others. However, preserving a balance of power is outside the scope of a paper focused on restraint. Restraint focuses on avoiding countries that have achieved a DSA from exploiting it, whereas a balance of power would prevent countries from acquiring a DSA in the first place. To some extent, this distinction is just a question of how to define a DSA: if a state has powerful AI that would by default give strategically significant capabilities but uses FlexHEGs to verifiably refrain from turning this AI into strategic capabilities, does this count as not acquiring a DSA or as demonstrating restraint with a DSA?}

\subsection{Institutional Design\label{sec:3.3-institutional-design}}

Finally, the U.S. could leverage its institutions to bolster the credibility of a commitment to restraint. We discuss four kinds of institutional design interventions.

\begin{enumerate}
\def\labelenumi{\arabic{enumi}.}
\item
  Appointing \textbf{personnel} with demonstrated commitments to the measured use of force could signal restrained preferences.
\item
  Codifying strict \textbf{internal usage processes} for AI systems could create friction against using them to trample adversaries.
\item
  Modifying \textbf{Special Access Programs (SAPs)} to increase transparency and oversight could reassure adversaries about how novel AI-derived capabilities would be governed.
\item
  Strengthening congressional \textbf{oversight of war powers} could increase the audience costs of reneging on restraint commitments.
\end{enumerate}

\subsubsection{Personnel\label{sec:3.3.1-personnel}}

The credibility of a commitment to restraint will depend on how adversaries perceive the individuals who might be making decisions to trample or to exercise restraint. Hence, the U.S. could strengthen a commitment to restraint by appointing people to roles in military or political leadership who have track records that are consistent with restraint.\footnote{Such a policy might be particularly important for roles that oversee Special Access Programs; these are DoD security protocols for safeguarding extremely sensitive information. We discuss Special Access Programs in more detail below. Per \cite{P-USDepartmentOfDefense2024g}, these roles include, at least, the USD(R\&E), USD(A\&S), USD(P), USD(I\&S) and the Director of the Special Access Program.} These track records could include, for example, statements or actions over the course of their career that demonstrate that they care deeply about the concept of state sovereignty. This mechanism might be weak on its own, because the personnel's commitment to restraint might not be visible -- both to those making domestic appointments and to international audiences -- but might be stronger when combined with public statements, diplomacy, or other institutional measures.\footnote{As before, there may also be a type shift problem: can states be confident that a given individual with a track record of restraint would still behave with restraint if the U.S. acquires powerful AI and so is in a very different strategic position to before?}

\subsubsection{Internal Usage Processes\label{sec:3.3.2-internal-usage-processes}}

By codifying how AI systems are used within government, the U.S. may be able to create additional signaling costs above and beyond model specs themselves. It could codify policies around model specs, including specific guarantees (as outlined above), monitoring and auditing if guarantees are ever removed, a principle of minimizing access to models that have had their guarantees removed, and public processes for changing the model spec. These signals may heighten reputational costs and create (minor, but nonzero) tying hands costs through frictions that would be required to use the model in ways that violate the public spec. 

Dedicating researcher hours and resources into the technical problem of ensuring that models follow a spec could also constitute an installment and sunk cost. Note, however, that the signaling value of those costs may depend on how much would be invested in the technical problem by default. Since getting models to follow a spec appears to be in the interests of both AI developers and governments \citep[][pp.~10--11]{P-Guest2023p}, dedicating researcher hours or resources to the technical problem may be of limited value absent a public proclamation of intended restraint.

\subsubsection{Special Access Programs\label{sec:3.3.3-special-access-programs}}

Novel weapons derived from powerful AI -- for example those built using superhuman R\&D agents or using novel dual-use technologies -- might be protected under a Special Access Program. SAPs are security protocols for safeguarding the Department of Defense's ``most sensitive classified information related to advanced systems, capabilities, technologies, and operations from adversary knowledge'' \citep{P-USDepartmentOfDefense2024g}. Because the novel military technologies that are likely to be most worrying to adversaries may fall under SAPs, changes to SAP procedures may serve as a signal of restraint. DoD could, for example:

\begin{enumerate}
\def\labelenumi{\arabic{enumi}.}
\item
  Amend SAP policy to bar ``waived SAP'' status -- which allows the Secretary of Defense to waive the inclusion of certain information in SAP reports and to share information only with the chair, ranking member, staff director, and designated security manager of the congressional defense and intelligence committees -- for any AI-derived programs;\footnote{Depending on how extreme the need to signal restraint becomes, a more forceful version of this suggestion could be to amend \cite{P-OfficeOfTheLawRevisionCounselUSHouseOfRepresentatives2025o} to \emph{forbid} waiver use for certain AI-derived weapon categories.}
\item
  Affirm that AI-derived weapons developed under SAPs would only be used to respond to aggression from adversaries;
\item
  Fold the Under Secretary of Defense for Policy (USD(P)) into diplomatic efforts as a principal with SAP oversight authorities.
\end{enumerate}

Each of these measures could signal restraint: increased oversight, public commitments, and peer-to-peer confidence building, respectively. Note, however, that they could also signal that novel AI-derived weapons are on the strategic horizon -- thereby exacerbating geopolitical competition on AI capabilities or creating some risk of sparking preventive conflict.\footnote{As before, the expected gains from restraint likely outweigh these risks because the signal is relatively weak evidence for the creation of a DSA, as opposed to some novel, but less powerful, military technology.}

\subsubsection{War Powers\label{sec:3.3.4-war-powers}}

The U.S. could attempt to signal its restraint by subjecting the decision to launch a first strike to greater congressional scrutiny and amending its public documentation concerning the use of force. We use ``first strike'' to mean the U.S. performing a preemptive attack that is designed to prevent or severely limit retaliation. Although the term is often used in the context of nuclear weapons policy, ``first strike'' in the sense that we are using it here need not involve nuclear weapons. Indeed, as discussed in the introduction, powerful AI may in the future equip countries with various nonnuclear but strategically significant capabilities.

While the approval of Congress is required to formally go to war via Article I of the Constitution, some scholars argue that Congressional authority over war powers has ``eroded over time'' \citep{P-Hathaway2024x}. In particular, the executive has sometimes invoked Article II of the Constitution to initiate ``limited'' force without seeking prior authorization from Congress, before reporting within 48 hours and then terminating hostilities within 60 days per the War Powers Resolution \citep{P-OfficeOfTheLawRevisionCounselUSHouseOfRepresentatives2025g,P-Weed2025z}. There is precedent for the claim that certain operations do not constitute ``hostilities'', such that the 60-day cessation provision does not apply \citep{P-Koh2011c} and that they do not ``rise to the level of war in the constitutional sense'' \citep{P-Engel2018j}, as in the cases of U.S. air strikes in Libya (2011) and Syria (2018), respectively. The position of the Trump Administration is similarly that congressional approval was not needed for the 2025 air strikes against the Iranian nuclear program, though some within Congress contest this position.\footnote{\cite{P-Weed2025n} The Iran strikes were partly justified as contributing to the self-defense of another nation. A similar logic could also be used in the context of powerful AI, with trampling justified as contributing to the defense of a U.S. partner or ally.}

Strengthening congressional authority over war powers could increase audience costs and reduce the risk of a unilateral decision to exploit a DSA. Several recent bills have aimed to revise or replace the War Powers Resolution to bring the judgment of ``both the President and Congress'' to bear when ``deciding whether the United States should engage in a significant armed conflict''.\footnote{See e.g. the National Security Reforms and Accountability Act \citep{P-UnitedStatesCongress2023e}, the National Security Powers Act of 2021 \citep{P-UnitedStatesCongress2021h} and the War Powers Consultation Act of 2014 \citep{P-UnitedStatesCongress2014k}. None were enacted, potentially due to the complexity of conflicting interests between the executive and legislative branches concerning the authority to go to war. Proposals that bundle war powers reform with other national security issues (as the National Security Powers Act did with arms export control and national emergencies) may face additional obstacles. Similarly, more ambitious proposals granting Congress a more authoritative role may be significantly less likely to pass than narrower consultation requirements.} New legislation could narrowly revisit the issue of war powers reform to ensure that Congress is consulted before the initiation of the use of force against a foreign state. Ensuring consultation, at a minimum, could codify an institutional opportunity for congressional objections to the exploitation of a DSA, thereby strengthening the credibility of tying hands signals. More ambitious proposals that give Congress a more authoritative role in the initiation of war could serve as even stronger signals \citep[e.g.][Sec.~4(a)]{P-UnitedStatesCongress2023e}.

Passing any form of legislation along these lines could signal Congress's intention to govern the use of military force; unilateral executive decisions to exploit a DSA that abrogate the will of Congress would be more likely to incur significant audience costs. But legislative language must be chosen carefully. The definition of ``significant armed conflict'' in the War Powers Consultation Act, for example, specifies ``any combat operation involving members of the armed forces lasting more than a week or expected by the President to last more than a week''. A first strike that employs novel AI-derived technologies could arguably fail both of those conditions: it could not ``involve'' human members of the armed forces at all or be expected by the president to last significantly less time than a week.

Beyond public initiation of hostilities, a particularly salient avenue for exploiting a DSA is \emph{covert action}, defined in Title 50 of the U.S. Code as ``an activity or activities of the United States Government to influence political, economic, or military conditions abroad, where it is intended that the role of the United States will not be apparent or acknowledged publicly'' \citep{P-USCode2023t50s3093,P-Hathaway2021t}. By default, per 50 U.S.C., Sec. 3093(a) and (c)(1), the president must notify congressional intelligence committees ``as soon as possible, but not later than 48 hours'' after the approval but before the initiation of a covert action. There are, however, three carve-outs that limit oversight:

\begin{enumerate}
\def\labelenumi{\arabic{enumi}.}
\item
  Under ``extraordinary circumstances'', reporting can be limited to the so-called ``Gang of Eight'', rather than the broader intelligence committees.
\item
  If a finding is \emph{not} reported, the president shall ``fully inform the congressional intelligence committees in a timely fashion and shall provide a statement of the reasons for not giving prior notice'' \citep[Sec. 3093(c)(3)]{P-USCode2023t50s3093}.\footnote{\citet{P-Rosenbach2009f} note that Congress and the White House have ``disagreed on the meaning of the requirement for reporting in a `timely fashion'\,'', which has ``led to friction between the two branches in the past''.}
\item
  After 180 days, the president can continue to withhold information from the broader intelligence committees by submitting a notification to the ``Gang of Eight'' that ``extraordinary circumstances affecting vital interests of the United States'' continue to hold \citep[Sec. 3093(c)(5)(B)]{P-USCode2023t50s3093}.
\end{enumerate}

Each of these restrictions might cause Congress to be kept in the dark \emph{indefinitely}, significantly limiting the potential for Congress to oversee covert action. As a signal of restraint, Congress could tighten these restrictions by, for example, stipulating a duration in place of ``a timely fashion'', striking the 180-day carve-out altogether, or extending notice to armed services and foreign affairs committees where covert action plausibly constitutes the use of force. Such changes might bolster expected domestic audience costs by ensuring thorough and preemptive congressional oversight over potential first strikes conducted as covert action.

In terms of formal documentation, the DoD could update its autonomy directive (3000.09), codify certain principles into the specs of military systems that involve general intelligence, or establish a no-first-strike policy through, for example, a National Security Memorandum (NSM) or Standing Rules of Engagement (SROE).\footnote{AI-derived weapons may strictly fall outside the scope of DoD 3000.09, amending the directive may nevertheless constitute a significant public signal of restraint.} Performing these updates should be done carefully: updating the autonomy directive, in particular, could be construed as a signal that the U.S. government has information about novel AI capabilities, and so could accelerate capabilities competition between the U.S. and its adversaries.

More broadly, the FY2026 NDAA mandates the creation of an ``Artificial Intelligence Futures Steering Committee'' in the Department of Defense \citep{P-UnitedStatesCongress2025z}, the responsibilities of which include articulating ``ethical, policy, and technical guardrails'' to ``prevent the misuse of {[}.~.~.{]} advanced or general purpose artificial intelligence'' as part of the Department's strategy for adoption of AGI \autocite{P-NA2025z}. If enacted, this measure would create an opportunity to demonstrate restraint by reaffirming a commitment to respecting sovereignty or against conducting first strikes with military force derived from, or enabled by, powerful AI.

\section{Speculative Mechanisms\label{sec:4-speculative-mechanisms}}

The package of restraint measures described above may still be insufficient to make a commitment to restraint credible. In this section we explore more speculative mechanisms, which might allow the U.S. to commit to restraint in a more credible way. We focus on three such mechanisms. \textbf{Escrow mechanisms} would transfer valuable resources to third parties that the U.S. would forfeit if it violated restraint commitments. \textbf{Active shields} would build technical constraints into AI-derived weapons that physically prevent their use for first strikes. \textbf{Automatic degradation} would trigger punitive destruction or locking of key U.S. assets upon detection of a first strike.

We note that these mechanisms would likely require significant technical work to implement and that they lie far outside of the standard national security policy window. Today, it seems very unlikely that the U.S. would agree to give up its national sovereignty in such a fundamental way, and it is far from obvious that the U.S. should do so. Credibly demonstrating restraint would have to be (perceived as) much more valuable than it is today, and more research would also be needed to fully weigh the consequences of implementing these mechanisms. The investigations below are exploratory, sketching out mechanisms that could be used in a scenario where the risk of preventive conflict looms large and drastic measures might be necessary to achieve stability through restraint.

\subsection{Escrow Mechanisms\label{sec:4.1-escrow-mechanisms}}

An escrow is an arrangement where an independent third party holds assets on behalf of two or more contracting parties. The U.S. could transfer resources -- whether raw funds, assets, or potentially even computational resources -- to a third party and would only obtain those funds back on the condition that it does not violate certain commitments. In this way, an escrow mechanism embodies a reducible cost that could be arbitrarily scaled up to make a commitment to restraint credible. Escrow mechanisms, however, face at least three significant difficulties.

First, it is \textbf{unclear when the U.S. should receive the resources back from escrow}.\footnote{This is a challenge with reducible costs in general.} An arrangement in which the U.S. pays resources into an escrow account when it makes an initial commitment, and then sets a repayment schedule based on calendar time or on AI development thresholds may struggle to generate a credible signal; in principle, the U.S. could bide its time until it recuperates its resources from the escrow, and then engage in conflict directly afterward. One solution would be to identify a resource that the U.S. could place in escrow indefinitely while still retaining access to it for legitimate uses. Compute is a promising candidate: the U.S. could house a significant number of AI chips in a jurisdiction over which it has little control, while retaining the ability to access the chips remotely for inference or training. This arrangement seems well-placed to make indefinite escrow more palatable to the US. At the same time, were the U.S. to renege on a commitment to restraint, the jurisdiction housing the chips might be able to seize or destroy them as forfeit.\footnote{Harack et al. discuss a similar idea specifically in the context of verification and international agreements in \citet[][Sec. 3.6.1.]{P-Harack2025v}.} An alternative solution would be to set the repayment threshold as some significant amount of calendar time after the U.S. achieves a particular AI development threshold. This arrangement would give adversaries a finite time window during which the U.S. would be unlikely to trample them. This window could be used to stabilize the strategic situation, such as with negotiations about how to structure the international order or with adversaries being able to make progress on their domestic AI development, restoring some balance of power.

A second difficulty is \textbf{determining the amount of resources that are placed into escrow.} What, exactly, is the price of peace? Too large, and the amount is likely to strain political feasibility. Too low, and the amount is unlikely to deter an aggressive actor from exploiting a DSA. This problem is worsened by an implication of the power shift problem above. Because powerful AI promises transformative economic benefits, on top of the possibility of military prowess, politically feasible amounts of resources that are placed into escrow may do relatively little to shape the choice calculus of the US. An amount of resources that is very significant by today's standards might be trivial by the standards of a country that has acquired powerful AI.\footnote{Speculatively, one possible solution would be to place resources into escrow that would not become more abundant in a world with powerful AI. One example might be prized historical artifacts. Alternatively, the amount of resources placed into escrow could ramp up over time, as the U.S. economy grows.}

The third difficulty is that \textbf{an aggressive U.S. with powerful AI might be able to steal resources back from escrow.} For example, it might be able to use whatever military capabilities that it is using to trample adversaries to take the resources in escrow by force. A partial mitigation here is timing. The party in control of the escrow might be able to destroy the resources in escrow as soon as that party notices the U.S. trampling an adversary. As a result, the U.S. would need to target both the adversary and the escrow at the same time -- a harder task than targeting the adversary alone.

\subsection{Active Shields\label{sec:4.2-active-shields}}

Active shields are weapons systems with built-in constraints to limit or prevent their offensive use \citep{P-Drexler1987y}. Traditionally, as \citet{P-Drexler1987y} argues, ``defense has required weapons that are also useful for offense''. Advances in technology, however, might permit the design of systems that can sense and assess the presence of certain conditions before acting.\footnote{Since \citet{P-Drexler1987y}, the idea of creating technological constraints on the use of (autonomous) weapons has gained some traction. \citet{P-Arkin2009e} describes an ``Ethical Governor'', which checks behavior against certain constraints. \citet{P-Chavannes2020f} suggest that an ethical governor system could be used to ensure that autonomous weapons comply with international humanitarian law, but note that several technical challenges remain unsolved. Chief among them, as we argue below, is the challenge of operationalizing constraints into the signals that determine an active shield's decisions.} If a weapons system can \emph{only} fire in response to an adversary's aggression, then it \emph{cannot} be used to conduct a first strike.\footnote{There is some precedent to states constraining the ways in which weapons can be used. For example, permissive action links (PALs) are locks that states apply to nuclear weapons to prevent them from being fired without authorization \citep{P-OfficeOfTheDeputyAssistantSecretaryOfDefenseForNuclearMatters2020m,P-SteinFeaver1987}. PALs constrain whether an individual is able to launch the weapons, not whether the military as a whole is able to do so. Even so, PALs provide support for the idea that credibly locking certain weapons is feasible.} (As before, we use ``first strike'' to mean a preemptive attack that is designed to prevent or severely limit retaliation. Although the term is primarily used in nuclear policy, a first strike in the sense that we mean it here would not necessarily involve nuclear weapons.) Active shields would instantiate a \emph{foreclosure guarantee} that the U.S. could not use certain systems to conduct a first strike and trample its adversaries.\footnote{For some additional discussion in the context of hardware-enabled AI governance, see \citet[][pp.~33--34]{P-Aarne2025u}.} Note that a full guarantee might not be necessary to have strategically important effects: creating a situation in which a defensive strike would be likely to succeed, while an attempt to execute a first strike would be likely to fail, would still change an adversary's strategic calculus.\footnote{Drexler (correspondence). As we argue in the conclusion, changes to the strategic calculus that fall short of a full guarantee are more likely to be sufficient when we do not assume that the U.S. will obtain powerful AI or a DSA first and instead take into account the deep uncertainty of the present strategic situation.} Another class of systems would augment the standard constraints on system capabilities, such as range, with more granular constraints, such as geofencing.\footnote{Standard constraints involve blunt physical or design limitations, such as a missile's maximum range or a drone's operational radius. More granular constraints involve context-sensitive, presumably computationally-enforced, limitations. A geofenced Minuteman III could be programmed to automatically disable if it crosses into an adversary's airspace, even though the target lies within its maximum range.} Even if such weapons could be activated before observing an adversary's aggression, they would not be able to deliver their payload as long as the constraints hold.

Active shields would need to meet several criteria.\footnote{This discussion draws on Appendix B.2.4.2 of \citet{P-Garfinkel2024s}.} First, adversaries would need to be \textbf{confident that a given weapons system is indeed an active shield}. This might involve strong verification mechanisms and mechanisms to prevent the U.S. from tampering with a weapons system to remove the constraints on usage, as well as \textbf{mechanisms to check that the U.S. is not building replacement weapons systems that lack the active shield constraints}.

Second, active shields would need to constitute a \textbf{sufficiently large portion of the U.S. military arsenal}. If the U.S. had some active shields but also significant military capabilities that were not active shields, then adversaries might worry that the U.S. would simply use its remaining arsenal for offensive purposes.

Third, it would need to be \textbf{difficult for either the U.S. or U.S. adversaries to deceive the active shields} concerning whether the U.S. has been attacked. An active shield must involve some decision logic: taking in some input data about the world and assessing that data to check whether or not an arming condition is met. This decision logic must be robust against deception -- both by its operator and by adversaries -- in order to credibly deter conflict. (To the extent that the decision process is carried out by a machine learning model, this is the problem of \emph{adversarial robustness}.\footnote{For more on adversarial robustness, see \citet[][sec.~3.3.3]{P-Hendrycks2024c}. Recall, per Schelling, that credible deterrence involves two components: credible restraint, if an adversary cooperates, and credible threat, if an adversary defects.}) An operator might want to deceive the active shield, making the shield incorrectly believe that the operator had been attacked, so that the operator could use the shield to launch a first strike. Conversely, adversaries might want to trick the shields into thinking that an attack was not occurring, leaving the shields locked and the operator's defenses weakened.

States' incentives to ensure that they can trick the active shields might also make it harder to reach agreement on the shields' criteria. Each side has an incentive to advocate for conditions that it privately knows it can deceive. However, constraints that are narrower than ``no first strike'', such as geofencing, may be relatively robust to deception and easy to agree upon because they can be implemented with comparatively simple mechanisms. A no-first-strike constraint requires answering a complex classification problem, whereas a geofencing constraint demands answering ``where am I?'' The adversarial problem does not disappear entirely, but it might be easier to solve the narrower problem through credible verification mechanisms.

\subsection{Automatic Degradation\label{sec:4.3-automatic-degradation}}

The third speculative approach uses automatic degradation mechanisms. If credible evidence indicates that the U.S. performed a first strike, then key assets could be locked or degraded over some time horizon. The assets could, for example, be AI systems themselves, weapons, or financial infrastructure. A toy example would be a kill switch to the power supply of crucial military data centers that is automatically triggered if a U.S. first strike is detected.\footnote{Which may already exist in a conventional, nonautomatic form, due to advanced persistent threats like Volt Typhoon \citep{P-CybersecurityAndInfrastructureSecurityAgencyCISA2024g}.} This mechanism differs from both escrow mechanisms and active shields in subtle ways. Like an escrow mechanism, automatic degradation is a punishment: an automatic tying hands cost is incurred if the U.S. abrogates a commitment. In some sense, an automatic degradation mechanism can be seen as an escrow mechanism where the third party that ``controls'' an asset is the degradation mechanism itself.

An automatic degradation mechanism could use the same locking architecture as an active shield, discussed in the previous section, but with inverted signal/access logic: rather than locking out first strikes and triggering upon an adversary's attack, automatic degradation \emph{triggers} a lock (or destruction) upon its operator performing a first strike (or violating some other commitment).\footnote{This approach alters the adversarial robustness problem. In order to deceive an active shield and create a window for attack, an adversary would have to spoof the \emph{absence} of a signal. By contrast, deceiving an automated degradation mechanism requires spoofing the \emph{presence} of some signal. The technical difficulty of verifying the relevant signals might differ across the two cases. Absence-spoofing often reduces to availability jamming, which may be easier than fabricating a valid, authenticated message. Presence-spoofing may require authenticity subversion -- such as via forged sensor evidence or cryptographically valid but false triggers. Authenticity subversion may be harder than denial. For physical-sensing stacks, however -- such as those detecting radar, seismic data, or emissions -- false-positive deception may be easier than perfect suppression. Nevertheless, the adversarial robustness problems defined above, as well as the difficulty of obtaining agreement about what signals are appropriate triggers, will bite for automated degradation mechanisms.}

These mechanisms have some precedent, at least in structural terms. The \citet{P-UnitedNationsSecurityCouncil2015r} creates a somewhat automatic punishment. Any Joint Comprehensive Plan of Action (JCPOA) participant that notifies the Security Council of significant nonperformance starts a clock. Unless the Security Council adopts a resolution to continue sanctions relief, UN sanctions automatically reimpose. This snapback mechanism mirrors the structural form of an automated degradation mechanism: it is a self-executing punishment rule conditioned on an evidentiary signal. It is not clear, however, whether or not it would be desirable to include some form of short-circuit mechanism, analogous to the ability of the Security Council to adopt a resolution that halts the punishment. Such a mechanism may create excessive risk that a DSA-enabled state can circumvent the punishment; without such a mechanism, the system loses flexibility and potentially creates space for disastrous false positives.

Automatic degradation mechanisms involve two design choices: the duration and decay of the punishment. Key assets could be locked for some particular time horizon, rather than destroyed or locked indefinitely. While these parameters allow for costs to be decided at the negotiation table, they generate some of the same problems as do escrow mechanisms. What level of cost is sufficient to demonstrate credible restraint? This question is intertwined with the problems of verification and deception; the greater the proposed cost, the more certain the restrained party would want to be that the trigger signal cannot be manipulated by an adversary.

Automatic degradation mechanisms also face the succession problem discussed in \cref{sec:2.2-foreclosure-guarantees}. Even if existing AI systems or weapons are subject to degradation triggers, the U.S. could potentially train new AI systems without such constraints. A partial mitigation is that building alternative systems outside the degradation framework is itself costly and so degradation mechanisms provide some disincentive against circumvention even without additional safeguards. As a much more speculative mitigation, the degradation trigger could potentially also cover attempts to develop systems that lack these constraints: if the U.S. were detected building AI systems without automatic degradation mechanisms, this would itself trigger degradation of existing assets. This approach resembles the mechanism discussed in \cref{sec:2.2-foreclosure-guarantees}. As noted there, however, such locking-in mechanisms face significant technical challenges and carry significant risks of their own.

\section{Future research\label{sec:5-future-research}}

This paper has explored how the U.S. might commit to \emph{restraint,} i.e. how the U.S. could demonstrate to adversaries that it would not pose an existential threat to them, even if the U.S. possessed powerful AI systems that, by default, would allow it to easily trample them. Several open and underexplored questions about restraint remain. In this section, we sketch several such questions, in the hope of encouraging future research on the topic. We loosely group these research questions by field.

First, there are high-level strategic questions about restraint that might be well suited to scholars of international relations or other social sciences.

\begin{enumerate}
\def\labelenumi{\arabic{enumi}.}
\item
  \textbf{How can we ensure that credible commitments remain limited to reasonable demands, such as respecting sovereignty?} The ability for the U.S. to make commitments about how it would use powerful AI might help adversaries to coerce the U.S. into making particular commitments.\footnote{For a discussion of this problem in the context of using AI to facilitate cooperation, see \citet[][p.~31]{P-Dafoe2020t}.} Although respecting sovereignty might be a reasonable demand for the U.S. to accept, other demands might be less reasonable and not ones to which the U.S. should agree. Researchers should focus on commitment mechanisms that are specifically relevant to reasonable demands, such as sovereignty, or explore what kinds of demands would be reasonable.
\item
  \textbf{How can restraint serve as a bilateral, or multilateral, strategy?} This paper focuses on how the U.S. in particular could unilaterally, credibly commit to restraint. However, it might be valuable, for similar reasons, for other countries that might develop powerful AI to be able to commit to restraint. How would the ideas in this paper have to be adapted to them? Additionally, there may be stronger restraint mechanisms that are available if implemented as part of an agreement between multiple countries, as opposed to unilaterally. What would these mechanisms look like?
\item
  \textbf{How desirable is restraint compared to other strategies that might reduce the risk of preventive conflict?} As discussed throughout the paper, implementing commitments to restraint would come with various tradeoffs. How should we evaluate them? Do other strategies that might reduce risk of preventive war, such as deliberately preserving an international balance of AI capabilities \citep{P-Hendrycks2025p}, have more favorable tradeoffs?
\end{enumerate}

Second, even if the strategic dynamics around restraint were well understood and if demonstrating restraint seemed desirable, various policy design questions remain outstanding. The package of measures described in \ref{sec:3-a-sketch-of-implementing-restraint-measures}, for example, requires more detail-oriented policy research before being ready for implementation.

Third, implementing commitments to restraint, and, in particular, the more speculative mechanisms, would require progress on various open technical problems.\footnote{Progress on these problems might require expertise in machine learning or AI hardware.} We discuss three of them below and note that progress on these topics would also be valuable for reducing other risks associated with powerful AI.

\begin{itemize}
\item
  \textbf{Ensuring that AI systems reliably follow a model spec:} Writing sovereignty principles into model specifications could strengthen restraint commitments, but existing techniques like RLHF sometimes fail to ensure that AI systems demonstrate their intended behaviors. Creating more robust methods for ensuring compliance with model specs, particularly in high-stakes contexts where AI systems might be pressured to circumvent their specifications, would be valuable for implementing foreclosure guarantees around sovereignty. This problem overlaps substantially with broader work on AI alignment.\footnote{Alignment is defined in several ways, but often focuses on ensuring that AI systems behave in line with an articulated set of instructions, such as a model spec. See \citet[][sec. 3.i]{P-Gabriel2020g}.}
\item
  \textbf{Verifying claims about AI systems:} There are various open technical problems that would need to be solved to create verification regimes where countries demonstrate to each other that their AI systems comply with certain properties. For detailed overviews, see \citet{P-Baker2025k} and \citet{P-Harack2025v}. Making progress on these problems could be useful both for making commitments to restraint but also for making commitments or agreements about various other aspects of governing powerful AI. Verification is not just a technical problem and could also benefit from expertise in other domains, such as institutional design.
\item
  \textbf{Adversarial robustness:} Strong adversarial robustness would be needed for active shields, as discussed in \cref{sec:4-speculative-mechanisms}. However, contemporary machine learning systems are vulnerable to adversarial attack, and some experts expect this to remain the case without focused effort. This problem could linger even if AI systems become much more generally capable and progress is made on related topics, such as alignment \autocite{P-GleaveNAc}. Improvements in the adversarial robustness of AI systems would also be useful for several other safety and security risks associated with AI, but tackling the question in the specific context of active shields, or detecting military aggression, appears to be a neglected line of work in the public domain.
\end{itemize}

\section{Conclusion\label{sec:6-conclusion}}

We have discussed how the U.S. could credibly commit to restraint, i.e. to avoid trampling adversaries, even if the U.S. were to acquire powerful AI systems and an accompanying decisive strategic advantage.

In the most challenging scenario, where adversaries are very confident that the U.S. would acquire a DSA and would want to use this to trample them, a credible commitment to restraint seems challenging, though may be feasible with significant policy effort and technical breakthroughs. Under more relaxed -- and realistic -- assumptions, where adversaries may have these concerns but would retain significant uncertainties, restraint mechanisms could be particularly effective and valuable tools. Such mechanisms might keep adversaries sufficiently reassured that they remain below the threshold of concern needed to initiate preventive conflict.

The development of increasingly powerful AI systems will have major consequences for international security, including potentially increasing the risk of preventive war. Demonstrating restraint is a promising avenue for reducing such risks but has received little research effort thus far. We hope that this paper helps lay a foundation for research into restraint as a strategy for increasing stability in a world advancing toward powerful AI.

\appendix
\addcontentsline{toc}{section}{Appendices}

\section{Setup}
This appendix adapts the signaling games from \citet{P-Fearon1997g} and \citet{P-Powell2006u}. Those models do not include some of the characteristic aspects of strategic competition for powerful AI; nevertheless, the structure of these games allows us to draw out important structural aspects of the AI competition.

In Fearon's games, a defender signals their resolve -- their willingness to fight if challenged. In this way, the games we develop are an inverted version of Fearon's: State $A$ seeks to signal its restraint -- its willingness to refrain from using a DSA in ways that impugn the sovereignty of State $B$. State $A$ would aim to do so in hopes of avoiding preventive conflict, which could stem from State $B$ fearing that State $A$ lacks restraint and would impugn State $B$'s sovereignty upon obtaining a DSA. Despite this difference, the underlying models employ a similar sequential structure.

We explore a series of four games with different signaling mechanisms: tying hands, sunk costs, installment costs, and reducible costs. Because the games differ only in their payoffs, we lay out the structure of the game in this appendix before detailing the payoffs and solving the models in the following appendices.

The games involve one-sided incomplete information. Both sides know that exploiting a DSA to impugn State $B$'s sovereignty would be the worst outcome from the point of view of State $B$, while State $A$ has private information concerning its type -- whether or not it would prefer to violate State $B$'s sovereignty upon acquiring a DSA.

The game proceeds with the following timing:
\begin{quote}
At $t_0$, nature draws $\theta$, State $A$'s type, which is either 0 (restrained) or 1 (aggressive). State $A$ and State $B$ share a common prior $p(R) = \pi$. State $A$ then chooses a signal $m \geq 0$.

At $t_1$, State $B$ observes $m$ and chooses whether or not to engage in preventive conflict. If it does, the game ends; if it does not, then the game continues to $t_2$.

At $t_2$, State $A$ has developed powerful AI and has obtained a decisive strategic advantage. If it exploits the DSA, it receives a payoff that depends on its type and its signal, whereas if it does not, then the states receive a peaceful dividend, calibrated to be their respective zero points.
\end{quote}
\section{Tying Hands}\label{app:tying-hands}
If State $B$ chooses to engage in a preventive conflict, then the payoffs are $[-c, -c]$. Each side loses according to the cost of conflict. If State $B$ chooses not to engage in a conflict, then payoffs depend only on State $A$'s choice. If State $A$ chooses to exploit its DSA, then the payoffs are $[\theta(V_D) - m, -V_B]$, where $V_B > c$ and $V_D > 0$. State $A$ obtains gains from dominance according to its type -- a restrained type gains nothing over the share it would have gotten otherwise -- minus the cost of signaling, and State $B$ gets its worst-case outcome. Whereas if State $A$ does not exploit its DSA, then the payoffs are $[0, 0]$.\\\\
We begin by solving for a pooling equilibrium, in which:
\begin{quote}
At $t_0$, both State $A$ types send $m = 1$; at $t_1$, State $B$ does not engage in preventive conflict after $m = 1$ and does engage in preventive conflict after $m = 0$; at $t_2$, neither State $A$ type exploits its DSA.
\end{quote}
At $t_2$: $A_R$ weakly prefers not to exploit, since $0 \geq -m$; $A_A$ chooses not to exploit iff $0 \geq V_D - m$ iff $V_D \leq m$.

At $t_1$: if State $B$ observes $m = 1$, then it anticipates restraint, so it doesn't fight; if State $B$ observes $m = 0$, then it expects exploitation, so it fights.

At $t_0$: if a type deviates to $m = 0$, State $B$ fights, and State $A$ obtains $-c$; signaling and obtaining status quo payoffs guarantees better outcomes, and so no one deviates.

Hence, for the tying hands model, a pooling PBE exists if, and only if, $V_D \leq m$.

This model has no separating equilibria. In other words, there is no equilibrium in which:
\begin{quote}
At $t_0$, a restrained State $A$ sends $m = 1$, an aggressive State $A$ sends $m = 0$; at $t_1$, State $B$ does not engage in preventive conflict after $m = 1$, and after $m = 0$ State $B$ engages in preventive conflict; at $t_2$, a restrained State $A$ does not exploit its DSA, while an aggressive State $A$ does so.
\end{quote}
The non-existence of a separating equilibrium follows from the ordinal preference orderings.

For State $B$: $(\text{not exploit} \mid \text{not fight}) \succ (\text{fight}) \succ (\text{exploit} \mid \text{not fight})$.

For a restrained State $A$: $(\text{not exploit} \mid \text{not fight}) \sim (\text{exploit} \mid \text{not fight}) \succ (\text{fight})$. 

And for an aggressive State $A$: $(\text{exploit} \mid \text{not fight}) \succ (\text{not exploit} \mid \text{not fight}) \succ (\text{fight})$.

Under these orderings, no Perfect Bayesian Equilibrium can separate such that $A_R$ sends a ``restraint'' signal $m$ that induces State $B$ not to engage in preventive conflict, while $A_A$ does not signal, inducing preventive conflict.
\begin{quote}
\textit{Proof (by contradiction).} Suppose such a PBE exists. Because $m$ is on-path for $A_R$, Bayes' rule pins State $B$'s beliefs so that after receiving $m$, State $B$ does not engage in preventive conflict. If $A_A$ deviates from its on-path behavior, signaling $m$, State $B$ still does not engage in preventive conflict. Once $A_A$ obtains a DSA, it chooses its preferred continuation, obtaining $\max\{V_D - m, 0\}$. Since $\max\{V_D - m, 0\} \geq 0 > -c$, $A_A$'s deviation strictly improves on the on-path conflict outcome. Thus we obtain a contradiction to $A_A$'s sequential rationality in the PBE. $\square$
\end{quote}
A separating equilibrium can occur, however, when the aggressive type -- but not the restrained type -- prefers an early preventive conflict to not exploiting a DSA. If State $A$ does not exploit its DSA, then State $B$ might develop its own powerful AI, which threatens to erase the DSA and lead to increasingly unstable competition. Presumably both State $A$ types place some credence on this possibility, which we can wind into the original payoffs. But it may be that an aggressive actor is also more concerned about this risk than a restrained actor.

We can bring this change into the model by modifying the payoffs if State $A$ does not exploit its DSA to $[-\theta(r), 0]$, where, as before, $\theta = 1$ if and only if State $A$ is aggressive, such that it is paranoid about the risk $r$.

Now we solve for a separating equilibrium at $m = m^*$:
\begin{quote}
At $t_2$: $A_R$ chooses not to exploit, since $0 > -m^*$; $A_A$ chooses not to exploit if and only if $-r > V_D - m^*$.

At $t_1$: after $m = m^*$, $P(\theta = R \mid m = m^*) = 1$, therefore no exploitation is anticipated, and so State $B$ doesn't engage in preventive conflict; after $m = 0$, $P(\theta = A \mid m = 0) = 1$, therefore State $B$ expects to be exploited, and so it chooses to engage in preventive conflict.

At $t_0$: for $A_R$, the on-path payoff is 0, while, if it deviates to $m = 0$, State $B$ engages in preventive conflict and State $A$ receives $-c$; hence, since $c > 0$ by construction, $A_R$ prefers $m = m^*$. For $A_A$, the on-path payoff is $-c$, while, if it deviates to $m = m^*$, State $B$ doesn't fight; at $t_2$, $A_A$ receives either $V_D - m^*$ or $-r$. Hence, to prefer $m = 0$, it must be the case that $-c \geq V_D - m^*$ and $-c \geq -r$, which implies $V_D \leq m^* - c$ and $c \leq r$.
\end{quote}
In intuitive terms, a separating equilibrium requires that the cost incurred by tying hands is sufficiently large that an aggressive type prefers to preventive conflict to exploiting a DSA after signaling and that an aggressive type prefers preventive conflict to leaving a DSA unexploited.
\section{Sunk Costs}\label{app:sunk-costs}
State $A$ can now pay a sunk cost $m$ that it incurs no matter what it does later on. If State $B$ chooses to engage in a preventive conflict, then the payoffs are $[-c - m, -c]$. If State $A$ chooses to exploit its DSA, then the payoffs are $[\theta(V_D) - m, -V_B]$, where $V_B > c$. If State $A$ does not exploit its DSA, then the payoffs are $[-m, 0]$.

Once again, there are no separating equilibria. As a result, there are no sunk cost solutions, because the sunk cost can generate no informational content. In other words, there is no equilibrium in which:
\begin{quote}
At $t_0$, a restrained State $A$ sends $m = 1$, an aggressive State $A$ sends $m = 0$; at $t_1$, State $B$ does not engage in preventive conflict after $m = 1$, and after $m = 0$ State $B$ engages in preventive conflict; at $t_2$, a restrained State $A$ does not exploit its DSA, while an aggressive State $A$ does so.
\end{quote}
As before, the non-existence of a separating equilibrium follows from the ordinal preference orderings. The argument proceeds as in Appendix~\ref{app:tying-hands} according to a proof by contradiction.

Unlike with tying hands signals, endowing the aggressive type with costs from not exploiting a DSA does not sustain a separating equilibrium. Let the payoffs from not exploiting a DSA be $[-\theta(r) - m, 0]$, where $r > 0$. Despite this change, there is no separating equilibrium.
\begin{quote}
\textit{Proof (by contradiction).} Suppose such a PBE exists. Because $m$ is on-path for $A_R$, Bayes' rule pins State $B$'s beliefs so that after receiving $m$, State $B$ does not engage in preventive conflict. If $A_A$ deviates from its on-path behavior, signaling $m$, State $B$ still does not engage in preventive conflict. Once $A_A$ obtains a DSA, it chooses its preferred continuation. Since $V_D - m > -r - m$, it will choose to exploit its DSA. Here, we proceed by cases:

\textit{Case 1:} $V_D - m > -c$, i.e., $m < V_D + c$. Hence, $A_A$ obtains $V_D - m$, and improves on its on-path payoff of $-c$. Thus, $A_A$ strictly prefers deviation, contradicting sequential rationality in the PBE.

\textit{Case 2:} $V_D - m < -c$, i.e., $m > V_D + c$. Hence, $A_A$ would not improve on its on-path payoff of $-c$. Now, however, $A_R$ would be faced with $0 - m$ from not exploiting its DSA or $-c$ from preventive conflict. But since $m > V_D + c$, we have $-m < -c$. And so the restrained type would also prefer conflict. Again, we obtain a contradiction to $A_R$'s sequential rationality in the PBE. $\square$
\end{quote}
As in the body of the text, the intuitive argument here is that $A_A$ faces a larger incentive to avert preventive conflict than does a restrained type; any signal that is sufficiently costly to make $A_A$ choose preventive conflict, and hence to yield a separating equilibrium, is sufficiently costly to make $A_R$ choose preventive conflict as well.
\section{Installment Costs}\label{app:installment-costs}
If State $B$ chooses to engage in a preventive conflict, then the payoffs are $[-c, -c]$. If State $A$ chooses to exploit its DSA, then the payoffs are $[\theta(V_D) - m, -V_B]$, where $V_B > c$. If State $A$ does not exploit its DSA, then the payoffs are $[-m, 0]$.\\\\
As in the sunk cost model, there are no separating equilibria. Because the installment cost is non-contingent -- borne whether or not State $A$ exploits its DSA -- it generates no informational content on its own. The argument proceeds as in Appendix~\ref{app:tying-hands} according to a proof by contradiction.

Endowing the aggressive type with risk-based costs when it does not exploit a DSA does not sustain a separating equilibrium. The argument proceeds as in Appendix~\ref{app:sunk-costs} according to a proof by contradiction.
\section{Reducible Costs}\label{app:reducible-costs}
If State $B$ chooses to engage in a preventive conflict, then the payoffs are $[-c - m, -c]$. If State $A$ chooses to exploit its DSA, then the payoffs are $[\theta(V_D) - m, -V_B]$, where $V_B > c$. If State $A$ does not exploit its DSA, then the payoffs are $[0, 0]$.\\\\
As demonstrated in Appendix~\ref{app:tying-hands} in the context of tying hands signals, the possibility of a separating equilibrium depends on how we model the aggressive state's payoffs. If the aggressive state only gains from the status quo -- with $V_D > 0$, with risk $r = 0$ -- then no separating equilibrium is possible. If, instead, the risk $r$ is larger than the expected cost of preventive conflict, then a separating equilibrium can be sustained. Both proofs proceed along the same lines as in Appendix~\ref{app:tying-hands}.

Pooling equilibria, by contrast, exist. Again, the conditions for pooling on restraint are:
\begin{quote}
At $t_0$, both State $A$ types send $m = m^*$; at $t_1$, State $B$ does not engage in preventive conflict after $m = m^*$ and, after $m = 0$, State $B$ engages in preventive conflict; at $t_2$, neither State $A$ type exploits its DSA.
\end{quote}
At $t_2$: $A_R$ prefers not to exploit, since $0 > -m^*$; $A_A$ chooses not to exploit iff $0 \geq V_D - m^*$ iff $V_D \leq m^*$. At $t_1$: if State $B$ observes $m = m^*$, then it anticipates restraint, so it doesn't fight; if State $B$ observes $m = 0$, then it expects exploitation, so it fights. At $t_0$: if a type deviates to any $m' < m^*$, State $B$ fights; because reducible costs appear in the conflict payoffs, any deviator's payoff is $-c - m' \leq -c < 0$; signaling and obtaining status quo payoffs guarantees better outcomes, and so no one deviates.

Hence, for the reducible costs model, a PBE where both types choose restraint exists if, and only if, $V_D \leq m^*$.
\section{Type Shifts}\label{app:type-shift}
In canonical signaling models, states are endowed with a type $\theta$ that is fixed throughout the model. The equilibria solutions in Appendices~\ref{app:tying-hands} through \ref{app:reducible-costs} depend on this type stability.

Yet, in practice, type is not stable. The relevant ``type'' is the degree to which State $A$'s decision-makers would perceive gains from exploiting its DSA. This perception can shift over time. Personnel changes can lead to new leadership with more aggressive preferences, and, even within a given administration, the aims of decision-makers can evolve. As a result, a signal that would be credible if type were fixed may fail to be credible if adversaries believe that type can drift -- and signaling may be even more difficult if adversaries believe that the type will drift towards aggression as a DSA approaches.

Formally, suppose that between $t_0$ and $t_2$ State $A$'s type evolves, such that there is some positive probability $p > 0$ that a restrained type becomes aggressive at $t_2$. In this case, the pooling equilibrium from Appendix~\ref{app:tying-hands}, which leverages tying hands signals, can break down.

At $t_2$, State $B$ no longer updates its beliefs after observing $m = 1$ to be $P(\theta = R \mid m = 1) = 1$. Instead, $P(\theta = R \mid m = 1) = (1 - p)$, because restrained types can evolve into aggressive types with probability $p$.\\\\
Anticipating this shift, State $B$ updates its decision at $t_1$. The expected payoff from not fighting becomes $(1 - p) \cdot 0 + p \cdot (-V_B) = -pV_B$. The payoff from fighting remains $-c$. Hence, State $B$ refrains from preventive conflict only if $-pV_B \geq -c$, i.e., $p \leq c/V_B$.

If $p$ exceeds $c/V_B$, then State $B$ is incentivized to choose preventive conflict even after observing the restraint signal $m = 1$.

This modeling choice is not the only way to depict the type shift problem; one could instead model the signaler's incentives as evolving stochastically and rederive incentive constraints for credible restraint.

Regardless of the precise modeling choice, the broader result is intuitive: the possibility of type shifts can undermine the credibility of signaling. Even if a signal is selected that would be credible \textit{ex ante}, the costs incurred by that signal may no longer be sufficient to bind decision-making when a DSA arrives.
\printbibliography

\end{document}